\documentclass[numberedappendix]{emulateapj}
\usepackage{apjfonts}
\usepackage{amsmath}
\usepackage{multirow}
\usepackage{array}
\usepackage[usenames]{color}
\allowdisplaybreaks[1]

\newcommand{\comment}[1]{}

\definecolor{purple}{RGB}{160,32,240}

\newcommand{\Msun}{M_{\odot}}

\newcommand{\plotadjust}[2]{\includegraphics[width=#2\columnwidth,type=eps,ext=.eps,read=.eps]{#1}}

\begin{document}

\shortauthors{BEHROOZI ET AL}
\shorttitle{Comparing Galaxies Across Time}
\title{Using Cumulative Number Densities to Compare Galaxies Across Cosmic Time}

\author{Peter S. Behroozi\altaffilmark{1}, Danilo Marchesini\altaffilmark{2}, Risa H. Wechsler\altaffilmark{1}, Adam Muzzin\altaffilmark{3}, Casey Papovich\altaffilmark{4}, Mauro Stefanon\altaffilmark{5}}
\altaffiltext{1}{Kavli Institute for Particle Astrophysics and Cosmology; Physics Department, Stanford University; 
Department of Particle Physics and Astrophysics, SLAC National  Accelerator Laboratory; 
Stanford, CA 94305}
\altaffiltext{2}{Department of Physics and Astronomy, Tufts University, Medford, MA 02155}
\altaffiltext{3}{Leiden Observatory, Leiden University, PO Box 9513, 2300 RA Leiden, The Netherlands}
\altaffiltext{4}{Department of Physics and Astronomy, Texas A\&M University,  College Station, TX 77843}
\altaffiltext{5}{Physics and Astronomy Department, University of Missouri, Columbia, MO 65211}

\begin{abstract}
Comparing galaxies across redshifts at fixed cumulative number density is a popular way to estimate the evolution of specific galaxy populations.  This method ignores scatter in mass accretion histories and galaxy-galaxy mergers, which can lead to errors when comparing galaxies over large redshift ranges ($\Delta z>1$).  We use abundance matching in the $\Lambda$CDM paradigm to estimate the median change in cumulative number density with redshift and provide a simple fit (+0.16 dex per unit $\Delta z$) for progenitors of $z=0$ galaxies.  We find that galaxy descendants do not evolve in the same way as galaxy progenitors, largely due to scatter in mass accretion histories.  We also provide estimates for the 1$\sigma$ range of cumulative number densities corresponding to galaxy progenitors and descendants.  Finally, we discuss some limits on cumulative number density comparisons, which arise due to difficulties measuring physical quantities (e.g., stellar mass) consistently across redshifts.  A public tool to calculate cumulative number density evolution for galaxies, as well as approximate halo masses, is available online. 
\end{abstract}


\section{Introduction}

\label{intro}

Galaxy surveys spanning a range of redshifts \citep[see, e.g.,][]{Grogin11,Coil11,Whitaker11,McCracken12,Muzzin13a} have allowed self-consistent studies of galaxy evolution over cosmic time.  Yet, comparing specific populations of galaxies across redshifts to determine the properties of their progenitors and descendants requires an assumption for how galaxies evolve.  An easy and popular approach is to compare galaxy properties at fixed cumulative number density over several redshifts (e.g., \citealt{Wake06,Tojeiro10,Brammer11,Papovich11,Tojeiro12,vDokkum13,Leja13} and references therein).  This approach ignores galaxy-galaxy mergers and scatter in mass accretion histories, which both affect the median cumulative number density of a galaxy population \citep{Leja13,Lin13,BWC12}.  Because individual galaxy merger and star formation histories are not well-constrained, more advanced comparisons have used either semi-analytical or semi-empirical galaxy-halo connections to infer galaxy evolution from simulated dark matter merger histories \citep[e.g.,][and references therein]{cw-08,Leitner11,BWC12,Moster12,Wang12,Lu12,Leja13,Lin13,Mutch13,Lu13}.

Previous work has addressed how cumulative number density changes for median progenitor \textit{or} descendant galaxies \citep{Leja13,Lin13}.  In this paper, we explicitly contrast progenitor and descendant galaxy evolution, and also address the significant scatter in progenitor and descendant galaxy cumulative number densities.  Here, we use abundance matching to identify galaxy cumulative number density with dark matter halo cumulative number density.  Using merger rates in dark matter simulations, we estimate the redshift evolution in the median and $1\sigma$ range in cumulative number density for any co-evolving galaxy population.  We discuss details of the method in \S \ref{s:method}, results in \S \ref{s:results}, interpretations in \S \ref{s:discussion}, and conclude in \S \ref{s:conclusions}.  In this work, we assume a flat, $\Lambda$CDM cosmology with parameters $\Omega_M = 0.27$, $\Omega_\Lambda = 0.73$, $h=0.7$, $n_s = 0.95$, and $\sigma_8 = 0.82$.

\section{Method}

\label{s:method}

\subsection{Abundance Matching Technique}

\label{s:tech}

To account for mergers and scatter in mass accretion histories, we use abundance matching between galaxies and dark matter halos in simulations \citep[see, e.g.,][]{Behroozi10,BWC13, BWC12, moster-09,Moster12, TG11, Yang11, Reddick12}.  Matching galaxies to halos means that halo merger trees can be converted into \textit{galaxy} merger trees (c.f.\ \citealt{Hopkins10}).  A full exploration of the information in these trees is beyond the scope of this paper, which we limit to the evolution of the most-massive or most-luminous progenitor and descendant galaxies.

Many ways exist to abundance match observed galaxies to dark matter halos in simulations \citep[see][for a review]{Reddick12}.  We match galaxies in rank order of decreasing stellar mass (or luminosity) to dark matter halos in rank order of decreasing peak historical halo mass.\footnote{Because dark matter is stripped from satellites more rapidly than stars, a halos' peak historical mass is a better proxy for the associated galaxy stellar mass than is its current mass\citep{Reddick12}.}  Abundance matching in this sense has been used successfully to reproduce galaxy clustering as a function of stellar mass or luminosity\footnote{Excluding bands which correlate more with the galaxy's star formation rate than its stellar mass, such as rest-frame B-band or UV.} and redshift, as well as galaxy conditional stellar mass functions \citep{conroy:06,moster-09,Reddick12,Watson13}.

Our technique can be summarized as follows:
\begin{enumerate}
\item Convert a galaxy cumulative number density at redshift $z_1$ to a halo mass with equal cumulative number density, using peak halo mass functions from \cite{BWC12}.
\item For halos at that mass at $z_1$, record the masses of the most-massive progenitor (or descendant) halos at $z_2$, according to the halos' mass accretion histories \citep{BehrooziTree}.
\item Convert the median halo progenitor/descendant mass at $z_2$, along with the 68$^\mathrm{th}$-percentile ($\sim1\sigma$) range of progenitor/descendant halo masses, into cumulative number densities using the halo mass function at redshift $z_2$.
\end{enumerate}

This method takes as input an initial cumulative number density, an initial redshift, and a final redshift.  The resulting final cumulative number density therefore does not depend on any additional properties of the initial galaxy population, such as stellar mass or luminosity.  That said, \textit{inferred} properties such as the change in the galaxies' stellar mass or luminosity will depend on the stellar mass or luminosity functions used (see \S \ref{s:limits}).

We provide a public implementation of this technique.\footnote{http://code.google.com/p/nd-redshift/}  As this process converts galaxy cumulative number densities into halo masses, our implementation also prints these out for convenience.  We currently do not consider scatter in stellar mass/luminosity at fixed halo mass; to the extent that merger rates per unit halo and specific mass accretion rates are weak functions of halo mass \citep{Fakhouri08,Fakhouri10,BWC12}, this scatter represents a second-order correction to the median cumulative number density evolution.  For the 68$^\mathrm{th}$-percentile range in progenitor/descendant cumulative number densities, this scatter may be more important, which is discussed in $\S \ref{s:limits}$.

\subsection{Dark Matter Simulation}

\label{s:dm_sim}

We use the \textit{Bolshoi} simulation \citep{Bolshoi}, which used the \textsc{art} code \citep{kravtsov_etal:97} to simulate a dark matter-only (250 $h^{-1}$ Mpc)$^3$ cosmological volume with 2048$^3$ ($\sim$8.5 billion) particles (1.73 $\times$ $10^{8}\Msun$ each).  The assumed cosmology was a flat, $\Lambda$CDM with parameters $\Omega_M = 0.27$, $\Omega_\Lambda = 0.73$, $h=0.7$, $n_s = 0.95$, and $\sigma_8 = 0.82$, similar to the WMAP9 best-fit cosmology \citep{wmap9}.\footnote{A future version of the code will adopt the Planck best-fit cosmology; however, cosmology dependencies in cumulative number density evolution are expected to be weak compared to the large scatter in mass accretion histories \citep{BWC12}.}  Halos were found with the \textsc{Rockstar} phase-space halo finder \citep{Rockstar}, and merger trees were generated with the \textsc{Consistent Trees} code \citep{BehrooziTree}.  From these merger trees, both halo mass accretion histories and merger rates were calculated \citep{BWC12}.

\section{Results}
\label{s:results}

We present results for the cumulative number density evolution for galaxy progenitors in \S \ref{s:progenitors}, for galaxy descendants in \S \ref{s:descendants}, a sample calculation of inferred stellar mass evolution in \S \ref{s:inferred}, and comparison with previous work in \S \ref{s:prev_work}.

\begin{figure}
\plotadjust{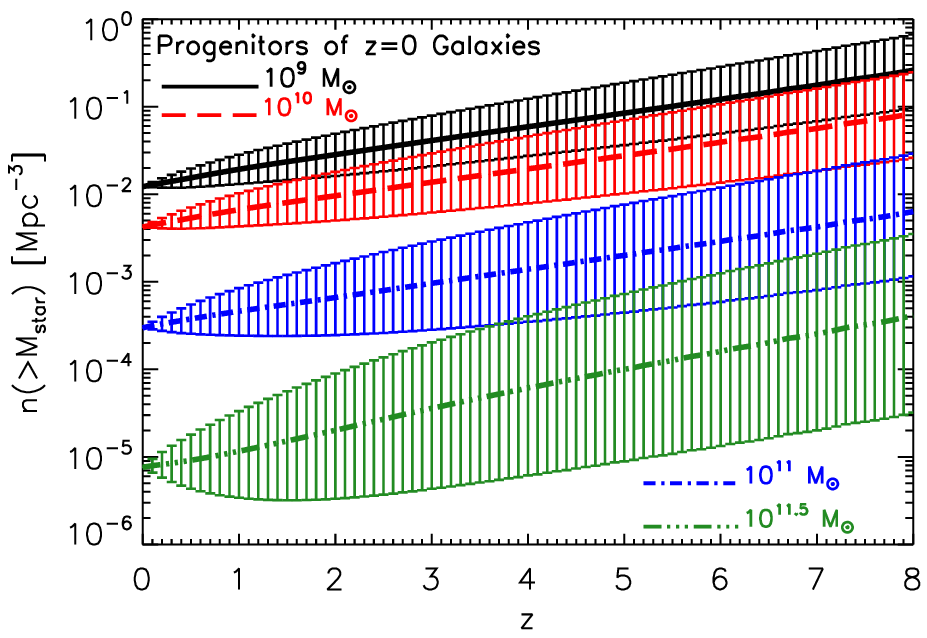}{1}
\caption{The evolution in median cumulative number density of the \textit{progenitors} of $10^{9}$, $10^{10}$, $10^{11}$, and $10^{11.5} \Msun$ galaxies at $z=0.1$.  The change in cumulative number density for the three lower-mass galaxies is excellently fit by $(0.16\Delta z)$ dex.  The 68$^\mathrm{th}$-percentile range of the corresponding progenitor cumulative number densities (error bars) grows more rapidly for larger halos.}
\label{f:nev}
\vspace{4ex}
\plotadjust{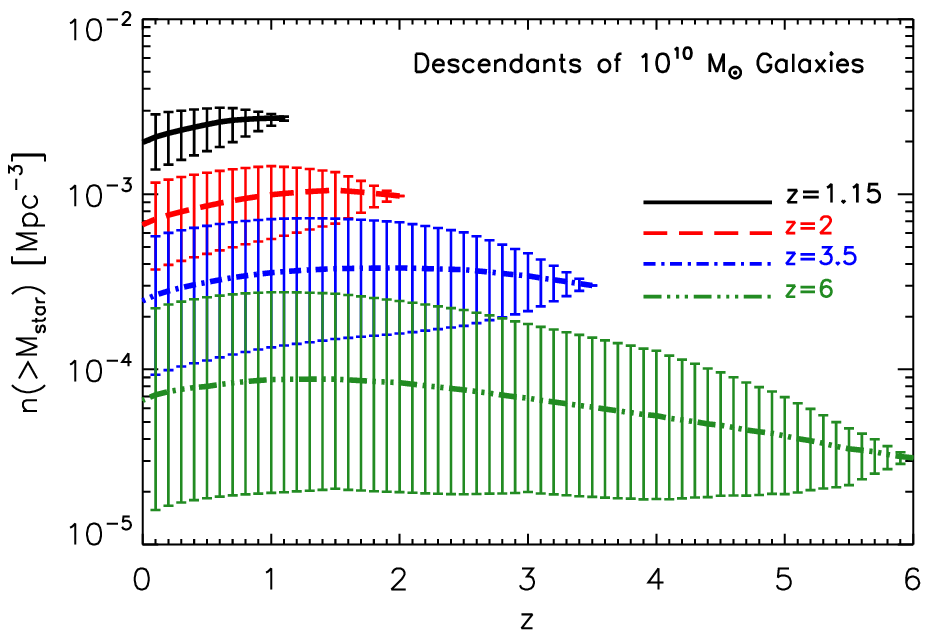}{1}
\caption{The evolution in median cumulative number density of the \textit{descendants} of $10^{10} \Msun$ galaxies at four separate starting redshifts ($z=1.15$, $2$, $3.5$, and $6$).  Contrast with the cumulative number density evolution of progenitors above.}
\label{f:nev_forwards}
\vspace{4ex}
\plotadjust{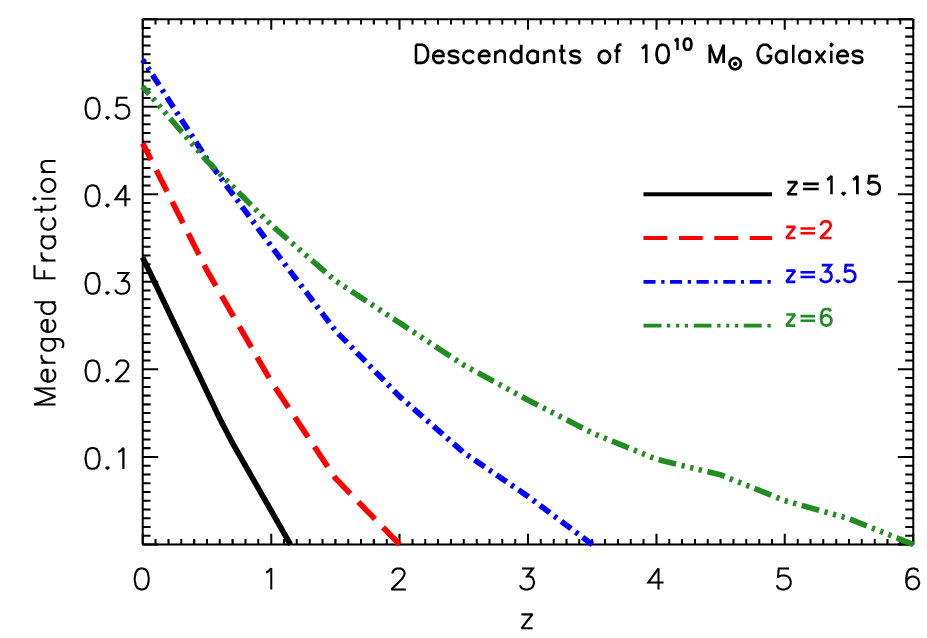}{1}
\caption{The fraction of descendants which have been lost due to mergers with a larger galaxy as a function of redshift, for $10^{10} \Msun$ galaxies at four separate starting redshifts ($z=1.15$, $2$, $3.5$, and $6$).}
\label{f:loss_forwards}
\end{figure}

\begin{figure*}
\begin{center}
\plotadjust{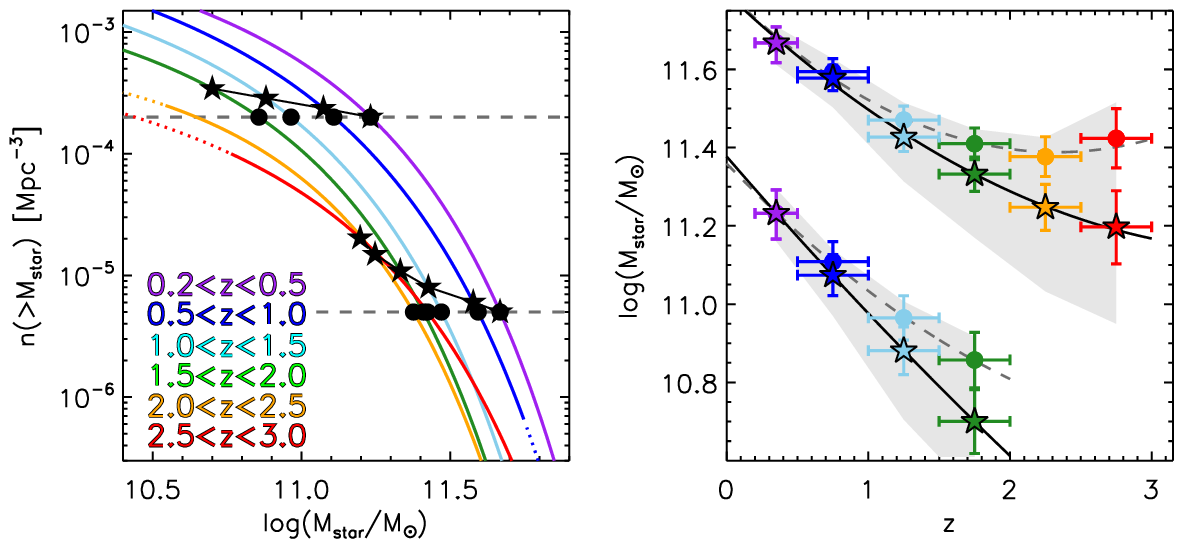}{1.5}
\end{center}
\caption{Example of galaxy progenitor stellar mass evolution inferred with and without the effects of mergers.  \textbf{Left panel:} Colored lines show integrated double Schechter fits to stellar mass functions from the UltraVISTA survey \citep{Muzzin13} for $0<z<3$.  Filled circles show the evolution in stellar mass for galaxies at fixed cumulative number densities of $5\times 10^{-6}$ Mpc$^{-3}$ and $2\times 10^{-4}$ Mpc$^{-3}$.  Filled stars show the difference if the evolving cumulative number density for galaxy progenitors (\S \ref{s:tech}) is included.   \textbf{Right panel:} \textit{Filled circles} correspond to those in the left panel (evolution in stellar mass for galaxies at a fixed cumulative number density); error bars show the width of the redshift bin as well as the formal uncertainty in stellar mass at a given cumulative number density from Poisson errors, sample variance, and photometric redshift uncertainties.  Including the effects of mergers and scatter in mass accretion histories (\textit{filled stars}) results in a 0.2 dex change in the inferred mass evolution of $10^{11.7}\Msun$ galaxies over this redshift range.  An even larger relative effect is seen for $10^{11.25}\Msun$ galaxies; despite the smaller redshift range over which they can be tracked ($z=0-2$ instead of $z=0-3$) in UltraVISTA, the change in inferred mass evolution is already $0.15$ dex.  The 68$^\mathrm{th}$-percentile ranges in cumulative number densities at $z\sim2$ are 1.5 dex and 1.3 dex for the $10^{11.7}\Msun$ and $10^{11.25}\Msun$ galaxy progenitors (respectively) at $z\sim 2$, corresponding to 68$^\mathrm{th}$-percentile stellar mass ranges (\textit{shaded regions}) of $\sim\pm 0.2$ dex for both.}
\label{f:ultravista}
\end{figure*}

\subsection{Galaxy Progenitors}

\label{s:progenitors}

In Fig.\ \ref{f:nev}, we show cumulative number density evolution tracks for the progenitors of $10^9$ to $10^{11.5}\Msun$ galaxies at $z=0.1$, calculated using the technique in \S \ref{s:tech}.  Cumulative number densities at $z=0.1$ were calculated from the stellar mass functions of \cite{Moustakas12}.  As discussed in \S \ref{s:tech}, the choice of stellar mass function only affects the initial cumulative number density for a given stellar mass and does not affect the cumulative number density evolution.

As shown in Fig.\ \ref{f:nev}, a power law describes the increase in cumulative number density towards higher redshifts for most galaxies; the change in cumulative number density is simply
\begin{equation}
(0.16 \Delta z )\textrm{ dex}
\end{equation}
as long as the galaxy's stellar mass is less than $\sim 10^{11.2}\Msun$ at $z=0$.  For $\sim 10^{11.5}\Msun$ galaxies at $z=0$, this rate increases to $\sim (0.22 \Delta z)$ dex (Fig.\ \ref{f:nev}).  For larger galaxies at $z=0$, we recommend use of the public tool in \S \ref{s:tech}, since the change in cumulative number densities is no longer well-fit by a simple power law.

The power-law functional form arises because merger rates are mostly constant per unit halo per unit $\Delta z$, regardless of time or halo mass, and have a largely power-law like dependence on mass ratio \citep{Fakhouri08,Fakhouri10,BWC12}.  More massive halos do have somewhat higher rates of major mergers \citep{Fakhouri08,Fakhouri10,BWC12}, which contributes to faster evolution in cumulative number densities.  Because the median halo mass as a function of stellar mass increases rapidly above $10^{11}\Msun$ in galaxy mass \citep{BWC12}, the effect is most prominent.  For lower stellar masses, the median halo mass changes more slowly, hiding the effect almost entirely.  We also note that a major merger in the history of a halo (i.e., a change of ~0.2-0.3 dex in its mass over a short time) will have the most impact on its corresponding cumulative number density for massive halos on the exponential tail of the mass function; these are also the halos which host $>10^{11}\Msun$ galaxies.  As such, for these halos and galaxies, relative rank ordering can change much more easily than for halos and galaxies of lower masses.

As discussed in \S \ref{s:limits}, the 68$^\mathrm{th}$-percentile range of progenitor cumulative number densities depends on how scatter in stellar mass at fixed halo mass changes with time.  Nonetheless, many qualitative features are robust.  For example, larger galaxies' progenitors extend over a larger range in cumulative number densities, as compared to smaller galaxies.  This is because a small change in stellar or halo mass equals a larger change in cumulative number density for more massive galaxies.  The large range in progenitor cumulative number densities can also result in significant differences between the median and the average progenitor mass.  For progenitors of $10^{11.5}\Msun$ galaxies, we predict a scatter of about $\pm$0.27 dex in stellar mass at $z=2.75$.  For log-normal scatter, this would imply that the median progenitor mass would be 0.08 dex less than the average progenitor mass.

Finally, we note that the progenitor cumulative number density ranges of $10^9$ and $10^{10}\Msun$ galaxies begin to overlap at $z=1.5$ in Fig.\ \ref{f:nev}.  Thus, it becomes difficult to tell which galaxies at $z=1.5$ become $10^{10}\Msun$ galaxies at $z=0$, and which become $10^{9}\Msun$ galaxies instead.  Current surveys are generally not deep enough for this to be a problem; e.g., progenitors of $10^{10} \Msun$ galaxies would be less than $10^{9}\Msun$ by $z=2$, below the completeness limit of most existing surveys \citep{BWC12}.

\subsection{Galaxy Descendants}

\label{s:descendants}

Fig.\ \ref{f:nev_forwards} shows cumulative number density evolution tracks for $10^{10}\Msun$ galaxies at a range of starting redshifts ($z = 1-6$).  Cumulative number densities at the starting redshifts were calculated from the best-fit model stellar mass functions in \cite{BWC12}.  At higher redshifts, $>10^{10}\Msun$ galaxies are rarer objects; at $z=6$, for example, they are typical progenitors of $10^{11.5}\Msun$ galaxies at $z=0$.

The median evolution in Fig.\ \ref{f:nev_forwards} is very different from Fig.\ \ref{f:nev}.  This is largely due to scatter in mass accretion histories and the shape of the halo mass function.  If one selects all progenitors at $z=z_1$ of halos with a given mass at $z=z_2$, many progenitor halos at $z_1$ will have typical accretion rates for their mass.  However, some fraction of halos at $z_1$ will always have unusually high accretion rates.  And, because smaller halos are always more numerous, small halos with high accretion rates will be disproportionately represented in the progenitor selection.  This selection effect explains qualitatively why cumulative number densities evolve more rapidly for progenitors than descendants (see also \citealt{Leja13}). 

Another important difference comes because of satellite galaxy mergers.  We show in Fig.\ \ref{f:loss_forwards} the fraction of galaxies which are lost due to mergers as a function of redshift for the same starting populations in Fig.\ \ref{f:nev_forwards}.  Every surviving galaxy at $z=0$ had a progenitor at all higher redshifts; however, a significant fraction of high-redshift galaxies never make it to $z=0$.  Because satellites cannot accrete matter easily, the likelihood that a galaxy merges has a strong correlation with the mass accretion history of its halo, and correspondingly with the cumulative number density rank of the halo.  Unfortunately, because of this strong correlation between mergers and halo rankings, it is difficult to give a clean theoretical interpretation of the shape of the cumulative number density evolution for galaxy descendants.

\subsection{Sample Calculation}

\label{s:inferred}

For a concrete example, we calculate the progenitor mass evolution of $10^{11.7}\Msun$ galaxies from the UltraVISTA survey \citep{Muzzin13} in Fig.\ \ref{f:ultravista}.  At $z=0$, these have a cumulative number density of $5\times 10^{-6}$ Mpc$^{-3}$.  Tracking their progenitors at fixed cumulative number density would imply that they had very little ($<0.1$ dex) mass evolution from $z=3$ to $z=1$.  Accounting for the effects of mergers gives a much more reasonable $0.2$-$0.25$ dex in mass growth from $z=3$ to $z=1$  \citep[see, e.g.,][]{BWC12}.

Fig.\ \ref{f:ultravista} also shows the progenitor mass evolution of $10^{11.25}\Msun$ galaxies.  In this case, the relative change in the median progenitor cumulative number densities is less than for the more massive $10^{11.7}\Msun$ galaxies.  However, due to the shallower slope of the stellar mass function at these masses, the relative change in \textit{stellar mass} is larger: 0.55 dex instead of 0.3 dex between $z=0$ and $z\sim2$.  For the same reason, the impact on inferred progenitor stellar masses from using fixed cumulative number densities instead of a more realistic evolving cumulative number density is more pronounced for $10^{11.25}\Msun$ galaxies as compared to the $10^{11.7}\Msun$ galaxies (Fig. \ref{f:ultravista}, right panel).

\subsection{Comparison with Previous Work}

\label{s:prev_work}

These results are in good agreement with those of \cite{Leja13}, who also find small changes in cumulative number density for descendants of semi-analytically modeled $z=3$ galaxies at $z=0$.  Using the cumulative stellar mass functions in \cite{Leja13}, our technique gives an implied median stellar mass evolution from $z=3$ to $z=0$ which differs on average by less than 0.05 dex from their reported values ($\sim0.6$ dex of growth for cumulative number densities between $5\times 10^{-5}$ and $8\times 10^{-4}$ Mpc$^{-3}$ at $z=3$).  These small differences could arise from their use of a semi-analytical model for galaxy formation, a fuller treatment of scatter in stellar mass at fixed halo mass in \cite{Leja13}, or a different model for satellite galaxy merger rates (see \citealt{Guo11}); we may explore these differences in more detail in future work.

\section{Discussion}

\label{s:discussion}

\subsection{Progenitors vs.\ Descendants}

As explained in \S \ref{s:descendants}, cumulative number density evolution is different for galaxy progenitors (i.e., backwards-looking comparisons) and galaxy descendants (i.e., forwards-looking comparisons).  Which direction to use depends on the targeted science question.  As an example, it is most relevant to use backwards-looking comparisons for galaxy star formation histories, as these can be directly compared to histories inferred from galaxy broad-band colors and spectra \citep[e.g.,][]{panter:06,Tojeiro09}.  For examining the fates of specific high-redshift populations at lower redshifts, as in analyzing the clustering evolution of luminous red galaxies \citep[e.g.,][]{White07,Wake08}, using a forward-looking comparison may be more appropriate.

\subsection{Limitations of Cumulative Number Density Comparisons}
\label{s:limits}

Luminosities can be measured fairly consistently across redshifts; however, the same is not necessarily true for stellar masses  \citep{Marchesini08,Conroy09,Muzzin09,Behroozi10,BWC12}.  Uncertainties arise because the appropriate priors for galaxies at one redshift may not be correct for galaxies at another redshift; these priors include star formation histories, dust content, the star-forming fraction of galaxies, metallicities, the initial mass function, fitting functions for galaxy light profiles, and many others \citep{Conroy09,Behroozi10,BWC12,Conroy12}.  Different pipelines applied to the same survey produce stellar mass functions which can differ in \textit{evolution} by up to $\sim 0.3$ dex in stellar mass \citep{BWC12}.  However, these uncertainties may well improve in the future.

The evolving quenched fraction \citep{Brammer11,Ilbert13,Muzzin13}, and the effects of dry stellar mass mergers \citep{BWC12,Moster12} will also affect this analysis.  Strictly speaking, directly converting stellar mass growth into star formation histories only applies for galaxy populations which are mostly star-forming across the entire redshift range considered and which have a low fraction of merger-deposited stellar mass.  Both these conditions are likely satisfied for, e.g., $\lesssim$ Milky-Way sized galaxies at $z=0$ \citep{Leitner11,BWC12}, and for UV-selected massive galaxies at high redshift \citep{Papovich11}.

Finally, the scatter in stellar mass at fixed halo mass will influence the inferred 1-$\sigma$ range of cumulative number densities for galaxy progenitors and descendants.  In this paper, we assume that the growth in stellar mass rank order differences with time is the same as the growth in halo mass rank order differences with time.  This will be the case in reality if individual galaxy star formation efficiencies depend much more on halo mass than on cosmic time or environment \citep{BWC13}.  Different assumptions---e.g., from semi-analytical or hydrodynamical galaxy formation models---may give different results for the scatter.

With these limitations kept in mind, cumulative number density comparisons provide a simple and valuable way to compare galaxies across cosmic time \citep{Brammer11,Papovich11,Lin13,Muzzin13,vDokkum13,Patel13}.

\section{Conclusions}

\label{s:conclusions}

We have presented a technique which robustly constrains the median cumulative number density evolution for galaxy populations and provides an estimate for the scatter in progenitor and descendant cumulative number densities (\S \ref{s:tech}).  
Our conclusions may be summarized as follows:
\begin{enumerate}
\item The evolution in cumulative number density for the most-massive progenitor galaxies is almost exactly  $(0.16\Delta z)$ dex for galaxies whose $z=0$ descendants have stellar mass $<10^{11.2}\Msun$ (\S \ref{s:progenitors}).  Ignoring this effect can lead to errors in inferred stellar mass evolution on the order of $0.2-0.3$ dex for a $\Delta z$ of 3 (\S \ref{s:inferred}).
\item For galaxy descendants, there is much less evolution in cumulative number density (\S \ref{s:descendants}).
\item For the exact evolution in the median and 1$\sigma$ range for galaxy cumulative number density evolution, as well as for calculating the corresponding halo mass at fixed cumulative number density and redshift, we provide a public tool at \texttt{http://code.google.com/p/nd-redshift/}
\item Galaxy cumulative number density comparisons across redshifts currently carry systematic errors in terms of the stellar mass evolution of $\sim 0.3$ dex (\S \ref{s:limits}).  Luminosity-based comparisons (e.g., for clustering evolution studies) suffer from fewer systematics, provided that the chosen luminosity band correlates more with galaxies' stellar masses than star formation rates.
\end{enumerate}

\acknowledgments
PB and RHW received support from an HST Theory grant; program number HST-AR-12159.01-A was provided through a grant from STScI, which is operated by AURA under NASA contract NAS5-26555.  DM acknowledges support from Tufts University Mellon Research Fellowship in Arts and Sciences. MS received partial support from NASA grants HST-GO-12286.11 and 12060.95.  We thank Ramin Skibba and the anonymous referee for helpful feedback.\\

\bibliography{master_bib}

\end{document}